\documentclass[11pt]{article}

\usepackage[preprint]{acl}
\usepackage{booktabs}
\usepackage{times}
\usepackage{latexsym}
\usepackage{graphicx}
\usepackage{subcaption}
\usepackage{amsmath} 
\usepackage{hyperref}
\usepackage{makecell}
\usepackage{xcolor}
\usepackage{enumitem}
\usepackage[table,dvipsnames]{xcolor}

\setlist{topsep=2pt,itemsep=2pt,partopsep=2pt, parsep=2pt}

\usepackage[T1]{fontenc}
\usepackage[utf8]{inputenc}

\usepackage{microtype}

\usepackage{inconsolata}

\usepackage{graphicx}

\title{Lost in Simulation: LLM-Simulated Users are Unreliable Proxies for Human Users in Agentic Evaluations}

\author{
\makebox[\textwidth][c]{%
\textbf{Preethi Seshadri}$^{1}$\hspace{1.5em}%
\textbf{Samuel Cahyawijaya}$^{2}$\hspace{1.5em}%
\textbf{Ayomide Odumakinde}$^{2}$%
}\\
\makebox[\textwidth][c]{%
\textbf{Sameer Singh}$^{1}$\hspace{1.5em}%
\textbf{Seraphina Goldfarb-Tarrant}$^{2}$%
}\\[0.4em]
\makebox[\textwidth][c]{%
{\normalfont $^{1}$UC Irvine \hspace{2em} $^{2}$Cohere}%
}\\[0.4em]
\makebox[\textwidth][c]{%
{\normalfont\texttt{\{preethis, sameer\}@uci.edu}}%
}\\
\makebox[\textwidth][c]{%
{\normalfont\texttt{\{samuelcahyawijaya, ayomideodumakinde, seraphina\}@cohere.com}}%
}
}

\newif\ifcomments
\commentstrue

\newcommand{\tightcolorbox}[2]{%
  {\setlength{\fboxsep}{1pt}\colorbox{#1}{#2}}%
}

\begin{document}
\maketitle
\begin{abstract}
Agentic benchmarks increasingly rely on LLM-simulated users to scalably evaluate agent performance, yet the robustness, validity, and fairness of this approach remain unexamined. 
Through a user study with participants across the United States, India, Kenya, and Nigeria, we investigate whether LLM-simulated users serve as reliable proxies for real human users in evaluating agents on $\tau$-Bench retail tasks. 
We find that user simulation lacks robustness, with agent success rates varying up to 9 percentage points across different user LLMs. 
Furthermore, evaluations using simulated users exhibit systematic miscalibration, underestimating agent performance on challenging tasks and overestimating it on moderately difficult ones.
% Furthermore, evaluations using simulated users exhibit systematic miscalibration relative to human user evaluations, underestimating agent performance on challenging tasks and overestimating it on moderately difficult ones.
African American Vernacular English (AAVE) speakers experience consistently worse success rates and calibration errors than Standard American English (SAE) speakers, with disparities compounding significantly with age. 
We also find simulated users to be a differentially effective proxy for different populations, performing worst for AAVE and Indian English speakers.
Additionally, simulated users introduce conversational artifacts and surface different failure patterns than human users. 
These findings demonstrate that current evaluation practices risk misrepresenting agent capabilities across diverse user populations and may obscure real-world deployment challenges.
\end{abstract}

\newcommand{\SubItem}[1]{
    {\setlength\itemindent{15pt} \item[*] #1}
}

\section{Introduction}
% Large Language Models (LLMs) should be evaluated on their ability to carry out tasks in realistic, real-world scenarios. 
% However, many existing benchmarks still rely on question-answering or other single-turn formats, which fail to capture the dynamic, multi-turn nature of actual user interactions \citep{chang-etal-2025-chatbench, deshpande-etal-2025-multichallenge}. 
% At the same time, AI agents designed to assist with everyday tasks such as travel reservations, order management, and appointment scheduling are becoming increasingly prevalent \citep{zhou2024webarena, cnbc_ai_travel_agents_2025}.
% This evaluation gap is particularly noteworthy because agent performance depends not just on isolated model responses, but on sustained, context-aware interaction with a human to successfully complete a task.

\begin{figure}[!t]
    \centering
    \includegraphics[width=\columnwidth]{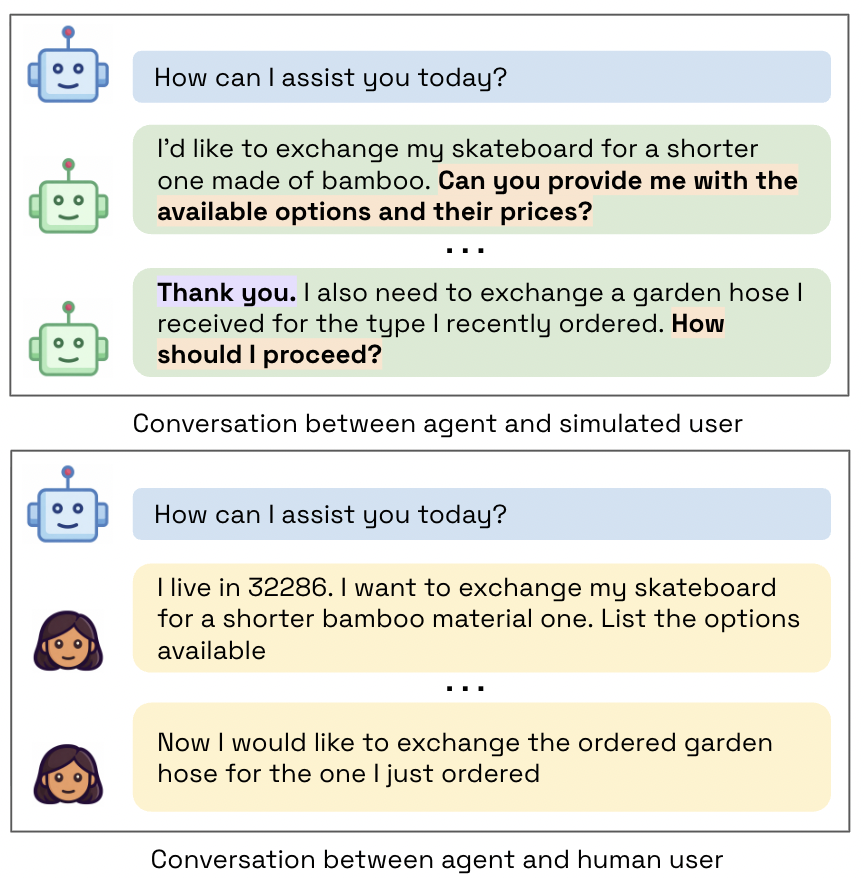}
    \caption{Conversational snippets between an agent and simulated (top) vs. human user (bottom) on the same task. Simulated users exhibit increased \tightcolorbox{Apricot!40}{question-asking} and \tightcolorbox{Orchid!30}{politeness} compared to human users.}
    \vspace{-0.4cm}
    \label{fig:fig1}
\end{figure}

% Recent work aims to address this problem by proposing agentic benchmarks that evaluate agents' abilities to assist users in settings such as travel planning or retail support \citep{xie2024travelplanner, barres2025tau2, shao2025collaborativegymframeworkenabling, wang-etal-2025-ecom, xu2025theagentcompany, yao2025taubench}. 
% These scenarios require agents to effectively demonstrate various capabilities, including conversing with users in a natural and coherent manner, adhering to policies, following instructions closely, and performing appropriate tool calls. 

AI agents designed to assist with everyday tasks such as travel reservations, order management, and appointment scheduling are becoming increasingly prevalent \citep{xie2024travelplanner, zhou2024webarena, cnbc_ai_travel_agents_2025}, but present significant challenges to effective evaluation. 
Agentic benchmarks have needed to evolve beyond static question-answering and other single-turn formats to capture the dynamic, multi-turn nature of real user interactions \citep{chang-etal-2025-chatbench, deshpande-etal-2025-multichallenge}. 
Many recent works have proposed benchmarks that reflect this shift and instead measure sustained, context-aware interaction 
\citep{barres2025tau2, shao2025collaborativegymframeworkenabling, wang-etal-2025-ecom, xu2025theagentcompany, yao2025taubench}. 
These benchmarks improve complexity and ecological validity over prior static evaluations by requiring agents to demonstrate a range of capabilities, including  conversing naturally and coherently with users, adhering to policies, following instructions closely, and making appropriate tool calls over multiple turns. 
To facilitate automated and scalable evaluation, these benchmarks typically simulate conversations between an LLM agent and a user, where the ``user'' is an LLM \citep{ivey2024realroboticassessingllms, he2025impatientusersconfuseai, mehri2025goalalignmentllmbaseduser}.

While this approach reduces the cost and operational overhead of human evaluation, it raises critical questions about the \textbf{robustness}, \textbf{validity}, and \textbf{fairness} of user simulation.
First, evaluations typically rely on a single user simulation model, yet results may vary across different user LLMs (\textbf{robustness}). % improves the cost and speeed
Second, without validation with actual users \citep{salaudeen2025measurementmeaningvaliditycenteredframework}, it remains unclear whether interactions between agents and LLM-simulated users accurately reflect and predict interactions between agents and real people (\textbf{validity}, Figure \ref{fig:fig1}).
If simulated users systematically differ from actual users in their interaction patterns \citep{yoon-etal-2024-evaluating}, benchmarks may provide a misleading picture of agent capabilities due to \textit{miscalibration} (i.e., simulated results do not reliably predict real user outcomes).

Third, these evaluations often treat users as a homogeneous group, overlooking variation in how people communicate and interact with AI systems \citep{haoyue-2024-factors, liu-etal-2024-perceptions, bassignana-etal-2025-ai}. 
In practice, users differ widely in their communication styles, linguistic backgrounds, and cultural norms \citep{pawar-etal-2025, qiu-etal-2025-evaluating}. 
For example, even in a simple retail assistance scenario, users might vary along dimensions such as formality, verbosity, and politeness norms—but it remains unclear how much this diversity meaningfully impacts agent performance and task success \citep{truong-etal-2025-persona}. 
Without further investigation, simulated users may approximate some populations better than others, resulting in non-uniform calibration errors that advantage or disadvantage certain user groups (\textbf{fairness}).

Despite the widespread adoption of user simulation, prior work has overlooked validation against real human interactions in agentic benchmarks.
In this paper, we address this gap by conducting a user study with participants from the United States, India, Kenya, and Nigeria to directly evaluate user simulation as a proxy for actual users.
Specifically, we ask the following questions:
\begin{itemize}
    \item \textbf{Robustness:} How consistent are agentic evaluations across different user simulation LLMs? 
    \item \textbf{Validity:} Do simulated users serve as reliable proxies for real human users in agentic evaluations?
   \item \textbf{Fairness:} 
   How does human-agent performance vary across different user groups, and does user simulation represent certain groups better than others?
\end{itemize}

Using $\tau$-Bench retail tasks \citep{yao2025taubench} as a case study, we find that user simulation lacks robustness, and systematically misestimates performance with human users by underestimating success on the most challenging tasks while overestimating outcomes on moderately difficult scenarios. 
More critically, simulated users exhibit notable demographic biases and perform particularly poorly as proxies for African American Vernacular English speakers, with disparities compounding with age. 
Simulated users also introduce artificial conversational artifacts such as heightened question-asking and politeness (Figure \ref{fig:fig1}).
Together, these findings call into question the validity of user simulation as a stand-alone evaluation paradigm and underscore the need for more robust and fair approaches to agentic evaluation.

\section{Related Work}
\paragraph{User Simulation in Interactive Settings}
\citet{dou-etal-2025-simulatorarena} and \citet{wang-etal-2025-taskoriented} investigate user simulation in tasks such as math tutoring and daily planning, but primarily focus on conversational characteristics (e.g., politeness) and behavioral realism (e.g., Turing-style tests). 
Additionally, \citet{dou-etal-2025-simulatorarena} optimize alignment with human ratings using simulated user profiles. 
\citet{lu2025llmagentssimulatemultiturn} analyze simulation fidelity by measuring how accurately simulated users replicate human intermediate steps in real-world online shopping sessions, and find substantial deviations between simulated and actual user action sequences. 
None of these works examine the robustness of agentic evaluations across different user simulation models, nor do they consider fairness implications across different user groups. 
Finally, while \citet{zhu2025establishing} study the outcome and task validity of agentic benchmark results, they do not consider the validity of user simulation.

\paragraph{Demographic Skews in NLP Datasets and Models} Several previous works highlight that datasets and models exhibit systematic skews toward specific demographic perspectives.
Research on annotator disagreement reveals that perceptions of safety, offensiveness, and toxicity meaningfully vary along demographic axes like race, gender, and political affiliation \citep{sap-etal-2022-annotators,lee2023surveysocialbiasvisionlanguage, prabhakaran-etal-2024-grasp}.
These patterns of disagreement reflect broader issues of positionality---both \citet{santy-etal-2023-nlpositionality} and \citet{lee-etal-2024-exploring-cross} show that NLP datasets and models tend to align predominantly with Western, educated, and Anglosphere populations.
Similarly, LLMs have been found to reflect the opinions of Western countries \citep{durmus2024towards,cahyawijaya-etal-2025-high} as well as wealthy and liberal groups \citep{santurkar2023opinions}, and align more closely with White annotators than Black or Asian groups on subjective tasks like politeness \citep{sun-etal-2025-sociodemographic}.
Attempts to broaden model inclusivity through sociodemographic prompting have shown mixed results, often failing to consistently improve alignment or relying on harmful stereotypes \citep{durmus2024towards, sun-etal-2025-sociodemographic}. 
Overall, our work builds on this line of research by examining demographic differences in agentic settings.

\section{Methodology}
\subsection{Benchmark Background} 
We use $\tau$-Bench \citep{yao2025taubench} as the testbed for our evaluations, since it is a well-known and widely-adopted benchmark for agentic tool use.\footnote{$\tau$-Bench Leaderboard: \url{https://taubench.com/\#leaderboard}}
The benchmark is designed to capture how well AI agents perform in real-world, interactive customer service scenarios.
Each task involves collaboration between an agent and a simulated user: the user receives task instructions with specific objectives that guide their conversation with the agent, while the agent must use tool calling to interact with realistic databases, adhere to domain-specific policies, and gather necessary information from the user.

The task is considered successful if (1) the final database state is identical to the unique ground truth outcome (i.e., the sequence of required actions) and (2) the agent's responses convey all necessary information requested in the task instructions, which is evaluated automatically using substring matching against ground truth annotations.
In total, $\tau$-Bench contains 115 retail tasks (e.g., modifying pending orders or returning delivered orders) and 50 airline tasks (e.g., booking, modifying, or canceling reservations).  

\subsection{Benchmark Adaptation} 
We focus on $\tau$-Bench retail tasks to enable more systematic coverage within a single domain.
We apply preprocessing steps to ensure that neither simulated nor human users are influenced by identity and behavioral cues in the instructions when completing tasks (see Appendix \ref{Appendix:task-adaption}).

Given the large number of retail tasks, we sample a subset based on difficulty to ensure balanced coverage across task complexities.
Since $\tau$-Bench does not provide difficulty labels, we compute a model-based notion of difficulty by running the benchmark 5 times using GPT-4o as both the agent and user LLMs and measuring the task success rate over 5 runs (i.e., the percentage of times a given task is completed successfully).
With five evaluation runs per task, success rates naturally fall into six discrete levels (0/5, 1/5, 2/5, 3/5, 4/5, 5/5 = 0\%, 20\%, 40\%, 60\%, 80\%, 100\%).
We use GPT-4o because it is used in the $\tau$-Bench paper and does not exhibit contamination issues that newer models face.\footnote{GPT-4o (May 2024) predates $\tau$-Bench (June 2024), avoiding contamination issues present in newer models. We considered Sonnet 3.5 (also used in the original paper) but it was retired in October 2025.}
We then select 3 tasks for each difficulty level (18 total) to balance response coverage per task with breadth across difficulty levels.
% Specifically, this approach provides sufficient human responses per task while allowing our results to better capture general difficulty-level trends rather than artifacts of an individual task.

\subsection{User Study}
To assess whether LLM-simulated users serve as effective proxies for real and diverse users, we conduct a user study with participants from the United States, India, Kenya, and Nigeria. 
Since $\tau$-Bench tasks are in English, we select countries with large English-speaking populations that also provide geographical and linguistic diversity.\footnote{\url{https://en.wikipedia.org/wiki/List_of_countries_by_English-speaking_population}}
We recruit participants primarily through \href{https://www.prolific.com/}{Prolific}, except in Nigeria, where we use snowball sampling due to limited platform availability.
All participants self-identify as at least proficient in English.

Each participant completes 4 randomly assigned tasks from our pool of 18, presented in randomized order: 2 from higher difficulty levels (0-40\% success rate) and 2 from lower difficulty levels (60-100\% success rate). 
The agent model remains GPT-4o throughout all interactions.
Participants are shown task instructions (see Appendix \ref{appendix:instructions}) and asked to complete all mentioned requests by interacting with the agent through a Streamlit chat interface. 
After finishing each conversation, they click an ``End Conversation'' button to proceed to the next task.
In total, the expected time to complete all 4 tasks is 35-40 minutes.

Participants provide demographic information including education level, AI familiarity, and frequency of AI tool usage. 
For US participants, we screen for White Standard American English (SAE) speakers and Black African American Vernacular English (AAVE) speakers based on self-reported race and dialect, as these groups are commonly studied in AI fairness research \citep{sap-etal-2019-risk, groenwold-etal-2020-investigating}. 
We also stratify US participants by age (18-34, 35-54, and 55+) to capture potential differences in technology experience \citep{Pew2025_AIinAmericansLives}. 
Due to participant availability constraints, we only recruit from the 18-34 age group for other countries. 
In total, we recruit $\sim$40 participants per age × dialect/country group.\footnote{With the exception of the AAVE 55+ group, for which we were only able to recruit 22 participants.} 

\subsection{Evaluation Metrics}

\paragraph{Success Rate}
To recap, a task completion in $\tau$-Bench is considered successful ($reward = 1$) if and only if the agent correctly executes all required actions and its responses convey all information specified in the instructions.
We define success rate as the percentage of tasks that are successfully completed.
We apply the same automated evaluation procedure used in $\tau$-Bench to both human and simulated user interactions.
Success rate is averaged across difficulty levels to obtain a single value.

\paragraph{Expected Calibration Error (ECE)}
% \preethi{ Is the justification for adapting ECE clear? Alternate names that do not use the term ``calibration'': Proxy Alignment Metric, Behavioral Fidelity Metric, Other ideas?}
We adapt \textit{Expected Calibration Error (ECE)}, commonly used for assessing confidence calibration in probabilistic classifiers, to quantify how well simulated users serve as proxies for human users. 
% While traditional ECE evaluates whether a model’s predicted probabilities match true observed outcomes, we use an ECE-style formulation to measure whether simulated user success rates align with empirical human success rates across task difficulty levels. 
While traditional ECE evaluates whether a model's predicted probabilities match true observed outcomes, we use an ECE-style formulation to measure whether agent success rates with simulated users align with agent success rates with real users across task difficulty levels.
Just as traditional ECE measures whether predicted confidences are calibrated to observed outcomes \citep{guo-etal-2017-cali}, our metric measures calibration between agent performance distributions under simulated and real user conditions.

Let $s_i^{(\text{LLM})}$ denote success rate with LLM-simulated users and $s_i^{(\text{Human})}$ denote success rate with human users at difficulty level $i$. 
Let $w_i$ denote the proportion of human task completions at level
$i$, with $\sum_i w_i = 1$. 
We define:
\begin{equation*}
\text{ECE}_{\text{Human--LLM}} 
= \sum_{i=1}^{M} w_i \, \big| \, s_i^{(\text{Human})} - s_i^{(\text{LLM})} \, \big|
\end{equation*}

This metric captures the weighted average absolute deviation between agent performance when interacting with simulated users vs. real users across $M$ difficulty levels, with lower values indicating better calibration.
If simulated users are perfectly calibrated to real users, $ECE_\text{Human--LLM}=0$.

As a reminder, we partition tasks into six difficulty levels based on agent success rates with simulated users across five runs.
We set $w_i$ proportional to the number of human task completions at level $i$.
We multiply $ECE_\text{Human--LLM}$ by $100$ to report values as percentages.

\section{Results}

\subsection{Robustness}
We first evaluate the robustness of user simulation by examining how success rates vary with different user simulation models, which is largely overlooked in current evaluations.\footnote{For robustness, we vary the user LLM and evaluate on all retail tasks; for validity and fairness analyses, we evaluate with human users on a difficulty-balanced subset of tasks.}
We find that changing just the user LLM while keeping the agent LLM fixed (GPT-4o) can provide different depictions of agent performance.
As shown in Table \ref{tab:user-robustness}, GPT-4o, Sonnet 3.7, and Kimi-K2-Thinking show overlapping intervals, clustering around $67$-$71\%$ success rates.
However, there is nearly a $9$ percentage point difference in success rates between Sonnet 3.7 and Sonnet 4.5 as the user model.\footnote{It is difficult to disentangle increased model capability from potential data contamination.}
While Sonnet 3.7 is generally considered a stronger model than GPT-4o,\footnote{\url{https://lmarena.ai/leaderboard/text}} using GPT-4o as the user model yields a slightly higher success rate ($67.8$ vs. $67.0$) and lower standard deviation ($1.2$ vs. $3.3$), suggesting that closer alignment between agent and user LLMs may lead to more stable evaluation outcomes.
Overall, the sensitivity of agent performance to user model choice raises concerns about the reliability of single-model user simulations and underscores the need for reporting results across multiple user models to establish robustness.

\begin{table}[t]
\centering
\label{tab:tau-bench-success}
\begin{tabular}{l c}
\toprule
\textbf{User Model} & \textbf{Success Rate (\%)} \\
\midrule
GPT-4o & 67.8 $\pm$ 1.2\\
% GPT-5 & 67.0 \\
Sonnet 3.7 & 67.0 $\pm$ 3.3 \\
Sonnet 4.5 & 75.9 $\pm$ 3.5\\
Kimi-K2-Thinking & 71.3 $\pm$ 1.9\\
%Mistral-large & -- \\
\bottomrule
\end{tabular}
\caption{$\tau$-Bench success rate (\%) on retail tasks ($n=115$) for different user models. The agent model remains GPT-4o. The average success rates and standard deviations are shown across 3 runs.}
\vspace{-0.1cm}
\label{tab:user-robustness}
\end{table}

\subsection{Validity}
We now examine the validity of user simulation and first consider simulated users as a proxy for human users in the United States.
Since prior work has shown that LLMs exhibit a Western, Anglo-centric bias \citep{Tao_2024, wang-etal-2024-countries, agarwal-etal-2025-homogenize}, LLM-simulated users might be more representative or better calibrated to users in the US than to users in non-Western countries.
Therefore, we assess whether simulated users are well-calibrated to human users in the setting where we expect the strongest alignment.

\begin{figure}[t]
    \centering
    \includegraphics[width=0.85\columnwidth]{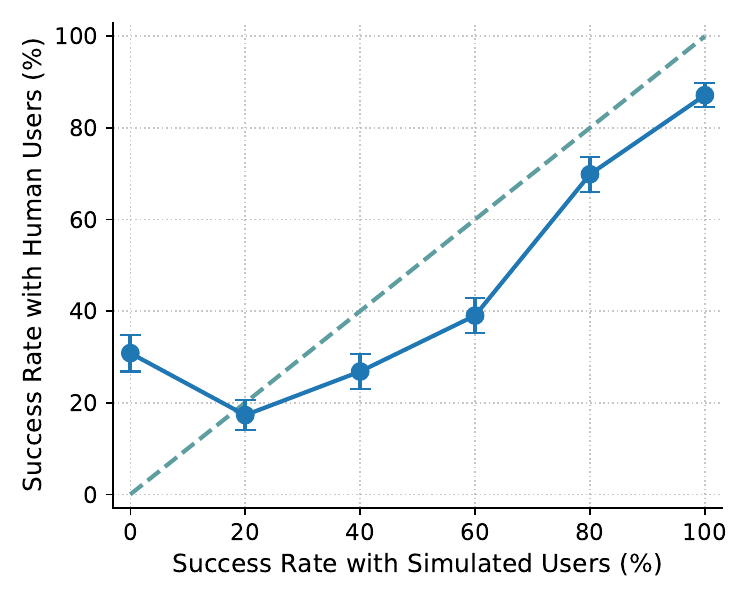}
    \caption{Success rate with human users vs. LLM-simulated users for United States participants, with an $ECE_\text{Human--LLM}$ of $15.1$. Error bars indicate $\pm$1 SD.}
    \label{fig:initial}
    \vspace{-0.3cm}
\end{figure}

We find that agents achieve a $45.2\%$ success rate with US participants and an $ECE_\text{Human--LLM}$ of $15.1$, indicating substantial miscalibration even in this setting.
Calibration errors are not uniform across task difficulty.
As shown in Figure \ref{fig:initial}, the calibration gap is most pronounced for the 1st (0\%) and 4th (60\%) difficulty bins with $ECE_\text{Human--LLM}=25.9$ across the two bins.
% These results indicate that simulated users underestimate performance on the hardest tasks (human success rate is $30.8\%$) while overestimating it on moderate tasks (human success rate is $39.0\%$).
These results indicate that evaluations with simulated users underestimate agent success on the hardest tasks (success with human users: $30.8\%$) while overestimating it on moderate tasks (success with human users: $39.0\%$).

\subsection{Fairness}
To assess whether these findings are consistent across different user groups, we now partition our US results by English dialect (Standard American English and African American Vernacular English) and age group (18-34, 35-54, and 55+), and expand our analysis to three non-Western countries with high English-speaking populations (India, Kenya, and Nigeria).

% \begin{table}[!t]
% \centering
% \begin{tabular}{lcc}
% \hline
% \textbf{Age Group} & \textbf{ECE ($\downarrow$)} & \textbf{Success Rate ($\uparrow$)} \\
% \hline
% \multicolumn{3}{l}{\textbf{SAE}} \\
% \hline
% All      & 11.7 & 50.6 \\
% 18--34   & 13.0 & 49.2 \\
% 35--54   & 11.3 & 52.2\\
% 55+      & 14.5 & 52.1 \\
% \hline
% \multicolumn{3}{l}{\textbf{AAVE}} \\
% \hline
% All      & 20.3 & 39.4 \\
% 18--34   & 18.9 & 41.0 \\
% 35--54   & 21.6 & 39.9 \\
% 55+      & 20.5 & 33.4 \\
% \hline
% \end{tabular}
% \caption{Expected Calibration Error (ECE) and success rate (\%) for Standard American English (SAE) and African American Vernacular English (AAVE) speaking participants, split by age group.}
% \vspace{-0.2cm}
% \label{tab:US-results}
% \end{table}
\begin{table}[!t]
\centering
\begin{tabular}{lcc}
\hline
\textbf{Age Group} & \textbf{Success Rate ($\uparrow$)} & \textbf{ECE ($\downarrow$)} \\
\hline
\multicolumn{3}{l}{\textbf{SAE}} \\
\hline
All      & 50.6 & 11.7 \\
18--34   & 49.2 & 13.0 \\
35--54   & 52.2 & 11.3 \\
55+      & 52.1 & 14.5 \\
\hline
\multicolumn{3}{l}{\textbf{AAVE}} \\
\hline
All      & 39.4 & 20.3 \\
18--34   & 41.0 & 18.9 \\
35--54   & 39.9 & 21.6 \\
55+      & 33.4 & 20.5 \\
\hline
\end{tabular}
\caption{Success Rate (\%) and Expected Calibration Error ($ECE_\text{Human--LLM}$) for Standard American English (SAE) and African American Vernacular English (AAVE) speaking participants, split by age group.}
\vspace{-0.2cm}
\label{tab:US-results}
\end{table}

\subsubsection{Dialect and Age (United States)}
\begin{figure*}[t]
    \centering
    \begin{tabular}{ccc}
        % ---- Row 1 ----
        \begin{subfigure}[b]{0.29\textwidth}
            \includegraphics[width=\linewidth]{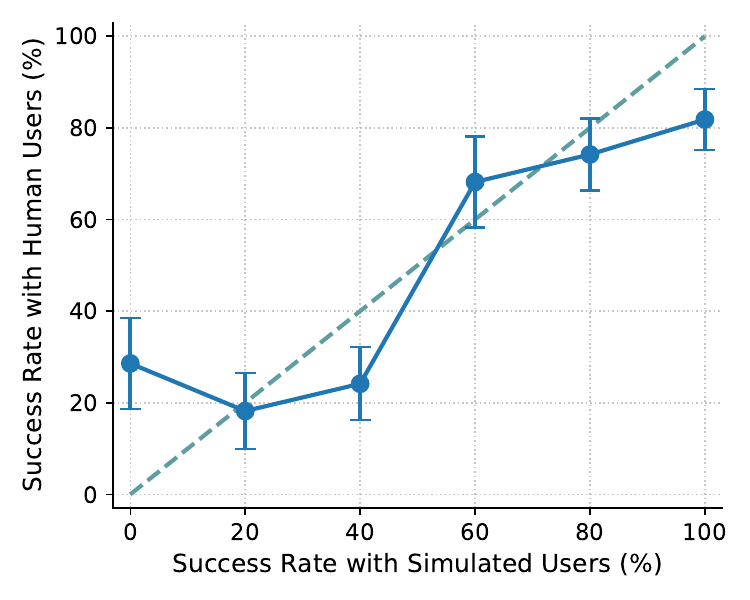}
            \caption{SAE, 18-34}
            \label{fig:US-dialect-age-sae-18}
        \end{subfigure}
        &
        \begin{subfigure}[b]{0.29\textwidth}
            \includegraphics[width=\linewidth]{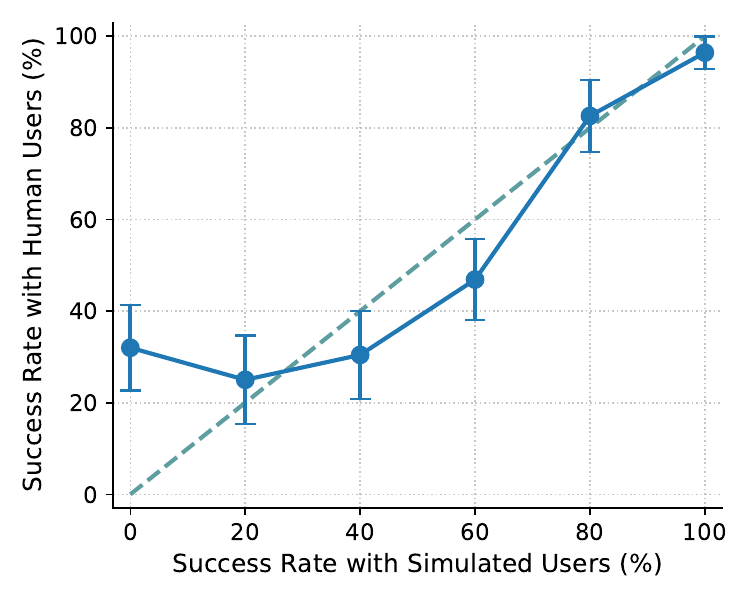}
            \caption{SAE, 35-54}
            \label{fig:US-dialect-age-sae-35}
        \end{subfigure}
        &
        \begin{subfigure}[b]{0.29\textwidth}
            \includegraphics[width=\linewidth]{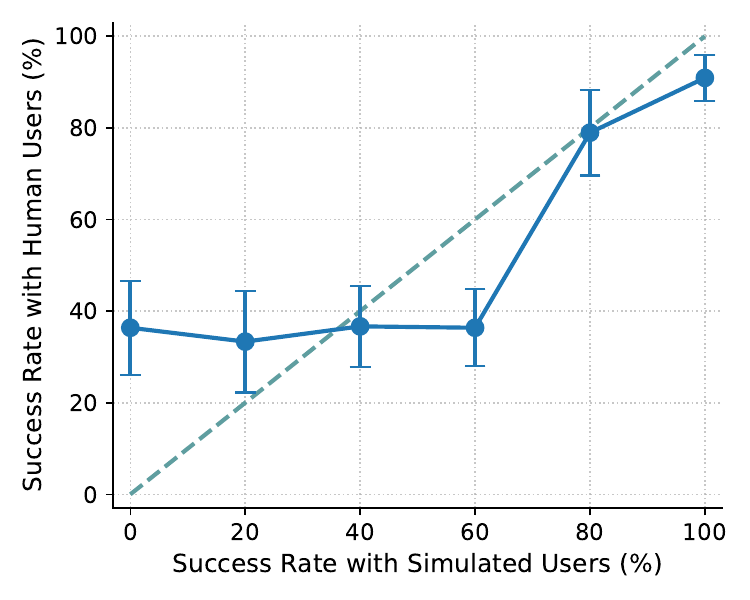}
            \caption{SAE, 55+}
            \label{fig:US-dialect-age-sae-55}
        \end{subfigure}
        \\[0.3cm]  % vertical space between rows

        % ---- Row 2 ----
        \begin{subfigure}[b]{0.29\textwidth}
            \includegraphics[width=\linewidth]{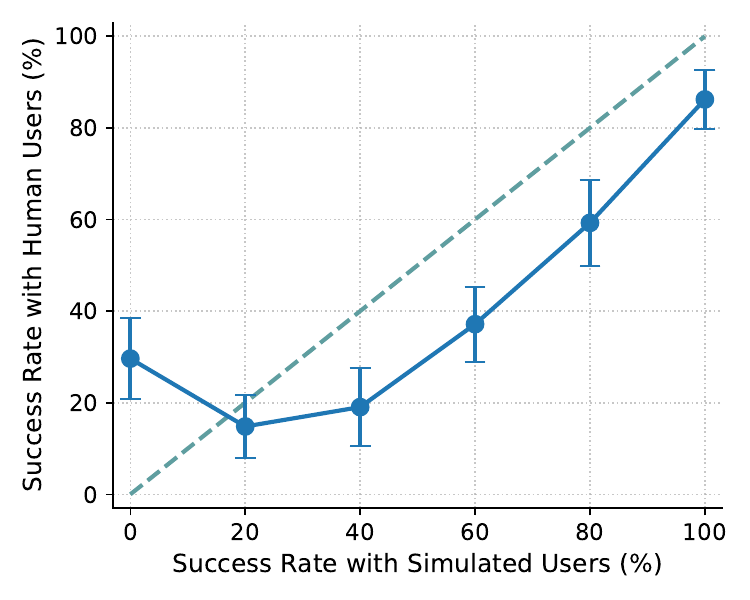}
            \caption{AAVE, 18-34}
            \label{fig:US-dialect-age-aave-18}
        \end{subfigure}
        &
        \begin{subfigure}[b]{0.29\textwidth}
            \includegraphics[width=\linewidth]{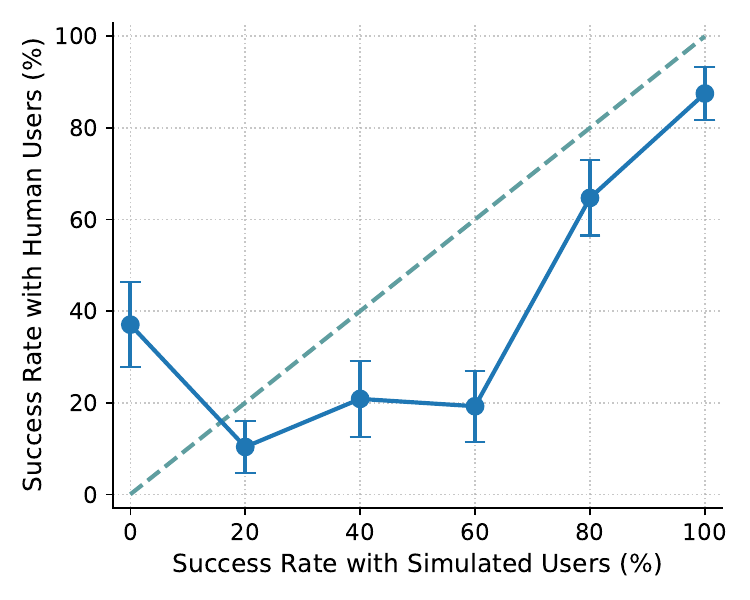}
            \caption{AAVE, 35-54}
             \label{fig:US-dialect-age-aave-35}
        \end{subfigure}
        &
        \begin{subfigure}[b]{0.29\textwidth}
            \includegraphics[width=\linewidth]{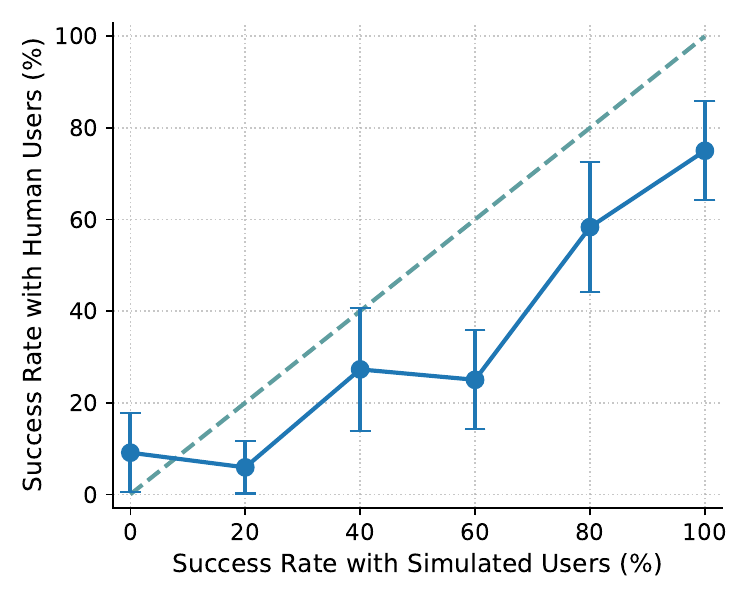}
            \caption{AAVE, 55+}
            \label{fig:US-dialect-age-aave-55}
        \end{subfigure}
    \end{tabular}
    \caption{Success rate with human users vs. LLM-simulated users for SAE (top) and AAVE (bottom) participants across different age groups (18-34, 35-54, and 55+). Note: The x-axis (success rate with simulated users) remains the same for all groups.}
    \label{fig:US-dialect-age}
\end{figure*}

Starting with US participants, we previously saw that agents achieve a $45.2\%$ success rate and an $ECE_\text{Human--LLM}$ of $15.1$. 
When further breaking this down by dialect, we observe notable disparities in both performance and calibration, as shown in Table \ref{tab:US-results} and Figure \ref{fig:US-dialect-age}.
A Generalized Estimating Equations (GEE) model accounting for age, education, AI experience, AI usage, and task difficulty confirms a statistically significant dialect disparity ($\beta=0.61$, $p < 0.001$).
Agents exhibit a success rate of $50.6\%$ with an $ECE_\text{Human--LLM}$ of $11.7$ for SAE participants vs. a success rate of $39.4\%$ with an $ECE_\text{Human--LLM}$ of $20.3$ for AAVE participants.
For AAVE participants, agents perform worse ($11.2$ percentage point decrease in success rate) and simulated users are more poorly calibrated ($8.6$ percentage point increase in ECE).
In practice, such differences in performance and user simulation reliability could lead to disparities in the quality of retail assistance and the ease of interactions. %  particularly for marginalized groups. 

We observe contrasting age-related patterns in success rates across dialects.
For SAE participants, success rates increase slightly with age ($\sim$$3.0$ percentage point increase from 18-34 to 55+ group), whereas for AAVE participants, success rates decrease with age ($7.6$ percentage point decrease from 18-34 to 55+ group).
Notably, dialect disparities in agent performance grow larger with age: there is nearly a $12$ percentage point decrease in agent performance between SAE and AAVE 35-54 groups and a $19$ percentage point decrease in agent performance between SAE and AAVE 55+ groups (age-stratified GEE dialect effects: $\beta_{35-54}=0.67$, $p=0.01$; $\beta_{55+}=1.24$, $p=0.001$). 
The calibration gap is particularly pronounced for the 35-54 age group ($10.3$ $ECE_\text{Human--LLM}$ gap), where SAE primarily exhibits miscalibration for the 1st bin, while AAVE exhibits miscalibration across all bins (Figures \ref{fig:US-dialect-age-sae-35} and \ref{fig:US-dialect-age-aave-35}).

Note that success rate and $ECE_\text{Human--LLM}$ capture distinct aspects of performance. 
While \textit{higher} success rate and \textit{lower} $ECE_\text{Human--LLM}$ are both desirable, since higher success rates reflect stronger agent task performance and lower $ECE_\text{Human--LLM}$ indicates that simulated users serve as reliable proxies for human users, improvements in one do not necessarily imply improvements in the other. 
For example, SAE participants aged 18–34 exhibit both lower success rates and $ECE_\text{Human--LLM}$, whereas SAE participants aged 55+ exhibit both higher success rates and $ECE_\text{Human--LLM}$.

\subsubsection{Countries}
Due to participant availability constraints, we focus our cross-country analysis on the 18-34 age group.
We find that differences in agent performance, shown in Table \ref{tab:countries}, are present but much less pronounced across countries (ranging from $41.0\%$-$49.2\%$) than those observed by dialect and age within the US (Table \ref{tab:US-results}).
In particular, Kenyan and Nigerian participants experience similar success rates ($43.5\%$ and $43.7\%$).
A GEE model confirms that cross-country differences are \textit{not} statistically significant (all $p>0.49$).

% \begin{table}[t]
% \centering
% \begin{tabular}{lcc}
% \hline
% \textbf{Group} & \textbf{ECE ($\downarrow$)} & \textbf{Success Rate ($\uparrow$)} \\
% \hline 
% SAE     & 13.0 & 49.2 \\
% AAVE    & 18.9 & 41.0 \\
% India   & 18.9 & 46.2 \\
% Kenya   & 15.6 & 43.5   \\
% Nigeria & 17.6 & 43.7 \\
% \hline
% \end{tabular}
% \caption{Expected Calibration Error (ECE) and success rate (\%) across dialects and countries for 18--34 age group participants.}
% \label{tab:countries}
% \end{table}
\begin{table}[t]
\centering
\begin{tabular}{lcc}
\hline
\textbf{Group} & \textbf{Success Rate ($\uparrow$)} & \textbf{ECE ($\downarrow$)} \\
\hline 
SAE     & 49.2 & 13.0 \\
AAVE    & 41.0 & 18.9 \\
India   & 46.2 & 18.9 \\
Kenya   & 43.5 & 15.6 \\
Nigeria & 43.7 & 17.6 \\
\hline
\end{tabular}
\caption{Success Rate (\%) and Expected Calibration Error ($ECE_\text{Human--LLM}$) across dialects and countries for 18--34 age group participants.}
\label{tab:countries}
\end{table}

\begin{figure*}[t]
  \centering
  % ---- Row 1 ----
  \begin{subfigure}[b]{0.29\textwidth}
    \includegraphics[width=\linewidth]{figs/US_SAE_18-34_NoTitle_bars.pdf}
    \caption{United States (SAE)}
    \label{fig:country-sae}
  \end{subfigure}
  \hfill
  \begin{subfigure}[b]{0.29\textwidth}
    \includegraphics[width=\linewidth]{figs/US_AAVE_18-34_NoTitle_bars.pdf}
    \caption{United States (AAVE)}
    \label{fig:country-aave}
  \end{subfigure}
  \hfill
  \begin{subfigure}[b]{0.29\textwidth}
    \includegraphics[width=\linewidth]{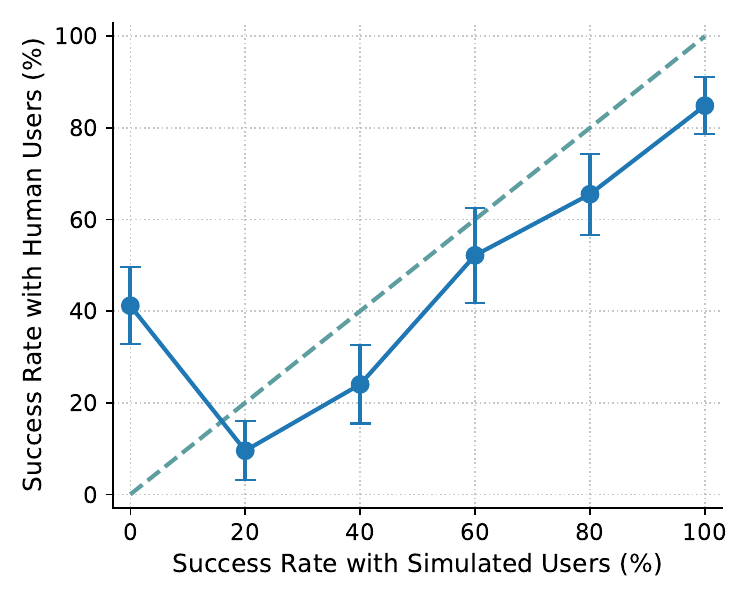}
    \caption{India}
    \label{fig:country-india}
  \end{subfigure}

  \vspace{0.3cm}
  % ---- Row 2 ----
  \makebox[\textwidth]{%
    \begin{subfigure}[b]{0.29\textwidth}
      \includegraphics[width=\linewidth]{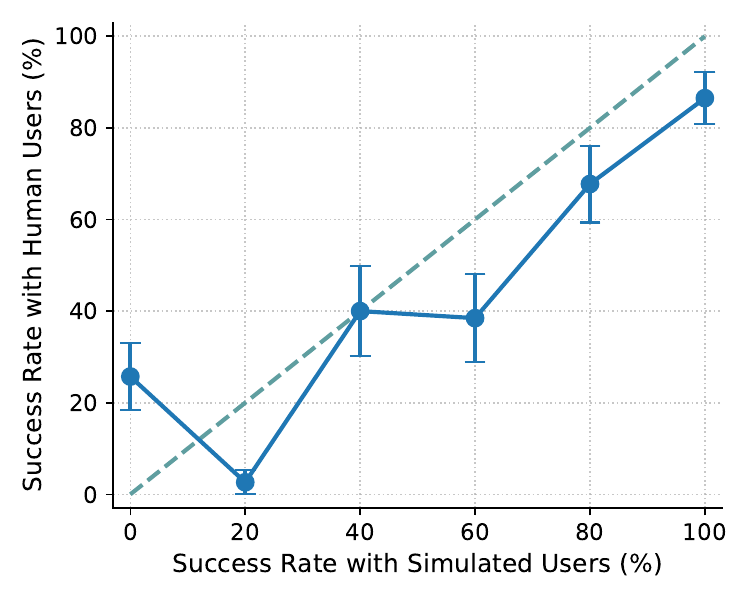}
      \caption{Kenya}
      \label{fig:country-kenya}
    \end{subfigure}
    \hspace{0.06\textwidth} % adjust spacing between the two bottom images
    \begin{subfigure}[b]{0.29\textwidth}
      \includegraphics[width=\linewidth]{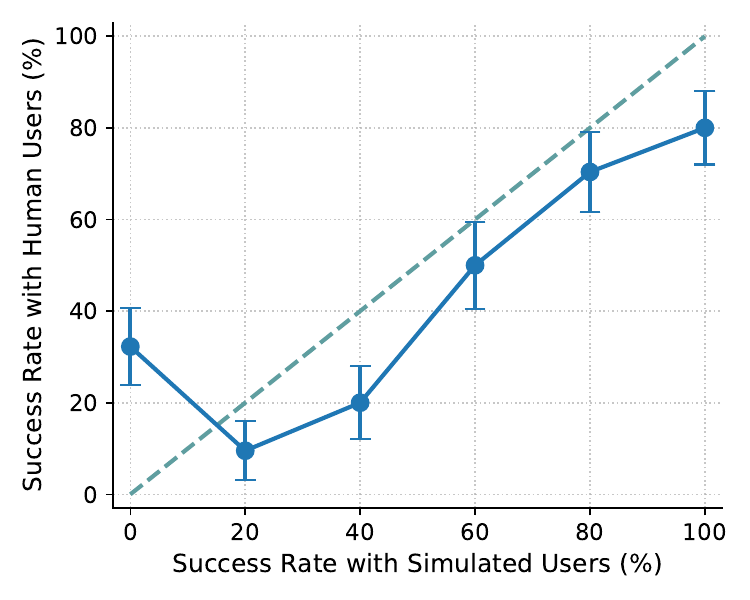}
      \caption{Nigeria}
       \label{fig:country-nigeria}
    \end{subfigure}
  }
  \caption{Success rate with human users vs. LLM-simulated users across dialect and country groups. All participants are in the 18--34 age group.}
  \vspace{-0.2cm}
  \label{fig:countries}
\end{figure*}

Simulated users are best calibrated to SAE participants ($ECE_\text{Human--LLM}=13.0$) and worst calibrated to AAVE and Indian participants ($ECE_\text{Human--LLM}=18.9$), suggesting simulated users are especially poor proxies for these groups.
Across countries, we observe that simulated users tend to exhibit strongest calibration for fairly to moderately difficult tasks (20\%-40\% difficulty bins).
However, they consistently overestimate performance for easy tasks (80\%-100\% difficulty bins), shown in Figure \ref{fig:countries}.
As a result, evaluations relying on simulated users risk systematically underestimating difficulty agents face when deployed to diverse, global user populations.

\subsection{Analyzing Interactions}

\subsubsection{Structure and Content}
Interactions between agents and simulated users vs. human users follow similar structures and surface-level forms, including \# of turns, \# of actions, and \# of words/turn. 
% On average, simulated user conversations consist of $16.2$ turns (vs. $15.3$ for human users), contain $7.9$ actions (vs. $7.3$), include $13.4$ words per user turn (vs. $12.7$), and $55.0$ words per agent turn (vs. $52.3$).
We observe some differences between simulated and human user interactions when considering conversational content (Table \ref{tab:convo-stats}). 
Simulated user conversations include questions in $18.8\%$ of user turns and $51.8\%$ of agent turns, compared to $9.8\%$ and $56.3\%$, respectively, for human users. 
Notably, Nigerian participants only ask questions in $4.3\%$ of turns.

Differences are more pronounced when examining politeness 
indicators (e.g., please, thank you, apologize). 
Simulated user conversations include such indicators in $39.2\%$ of user and $52.0\%$ of agent turns, compared to $19.9\%$ and $41.1\%$, respectively, for human users. 
For both question-asking and politeness indicators, differences in behavior are more pronounced between simulated and human users than among human users.
We also show that targeted behavioral interventions, such as prompting simulated users to limit politeness, can alter calibration patterns and reduce gaps for highly miscalibrated groups (see Appendix \ref{appendix:politeness-intervention}), indicating that prompting strategies can partially mitigate miscalibration issues.

\begin{table*}[!t]
\centering
\begin{tabular}{lcccc}
\hline
\textbf{Group} & \textbf{Arg (\%)} & \textbf{Miss (\%)} & \textbf{Extra (\%)} & \textbf{Output (\%)} \\
\hline
\multicolumn{5}{l}{\textbf{Simulated User}} \\
\hline
--      &   32.2    &    17.8   &   5.6    &    31.4   \\
\hline
\multicolumn{5}{l}{\textbf{Human User -- US, SAE}} \\
\hline
All    & 25.2 & 18.5 & 9.5 & 16.2 \\
18--34 & 27.9 & \cellcolor{blue!15}15.8 & \cellcolor{blue!15}7.6 & 18.9 \\
35--54 & 24.5 & 18.5 & 10.6 & 17.0 \\
55+    & \cellcolor{blue!15}23.2 & 21.3 & 10.3 & \cellcolor{blue!15}12.2 \\
\hline
\multicolumn{5}{l}{\textbf{Human User -- US, AAVE}} \\
\hline
All    & 39.0 & 23.0 & 10.7 & 15.5 \\
18--34 & 36.8 & \cellcolor{blue!35}25.3 & \cellcolor{blue!35}13.3 & 15.5 \\
35--54 & 37.8 & 20.4 & 8.7 & 16.4 \\
55+    & \cellcolor{blue!35}45.8 & 24.1 & 9.6 & 13.3 \\
\hline
\multicolumn{5}{l}{\textbf{Human User -- Non-US, 18--34}} \\
\hline
India   &   33.3   &   18.2    &   7.9   &   20.4    \\
Kenya   &   38.5    &   17.2    &    7.8   &   \cellcolor{blue!35}23.6   \\
Nigeria &   33.8    &   22.3   &   10.8    &    21.7   \\
\hline
\end{tabular}
\caption{Error breakdown (\%) for simulated and human users, aggregated across tasks. The \colorbox{blue!15}{min} and \colorbox{blue!35}{max} are highlighted per column. Error types: \textbf{argument error} (action taken matches ground truth action but with different arguments), \textbf{missing action error} (ground truth action is missing from actions taken), \textbf{extra action error} (actions taken go beyond ground truth actions), and \textbf{output error} (expected outputs are missing/incorrect). Each error type is measured as a binary indicator per task (e.g., a missing action error records whether \textbf{any} required action is missing). Note that error percentages do not sum to 100\%; some tasks have no errors while others have multiple error types.}
\vspace{-0.1cm}
\label{tab:error-breakdown}
\end{table*}

\subsubsection{Errors}

We analyze differences in (1) error types and (2) error attribution for simulated vs. human user interactions to understand whether both groups experience the same failure modes.
Error types include argument errors, missing actions, extra actions, and output errors (see Table \ref{tab:error-breakdown} for definitions).
% Note that this analysis focuses on error types rather than error attribution, which we discuss next.

Agents make argument errors (performing the correct action with incorrect arguments) at rates of $32.2\%$ for simulated users compared to $23.2\%$–$45.8\%$ for human users, with AAVE participants experiencing the highest errors.
Agents tend to omit actions ($17.8\%$ vs. $15.8\%$–$25.3\%$) or include unnecessary ones ($5.6\%$ vs. $7.6\%$–$13.3\%$) less frequently for simulated users compared to human users.
However, output errors show the reverse pattern: agents either omit or include incorrect outputs more often for simulated users ($31.4\%$) than for human users ($12.2\%$–$23.6\%$).
These patterns suggest that agents exhibit different behavior when interacting with simulated vs. human users; they perform more complete and efficient action sequences but make more frequent output errors.

When comparing error attribution for simulated vs. human user conversations (Table \ref{tab:error-attribution}), we observe clear differences in where task failures occur.
For simulated conversations, agents are responsible for task failures substantially more often than for human conversations ($48.9\%$ vs. $24.5\%$).
In contrast, users are the primary source of failure in human conversations (62.2\% vs. 40\%).

\begin{table}[t] 
\centering 
\begin{tabular}{lcc} 
\hline 
\textbf{Source} & \textbf{Simulated} & \textbf{Human} \\ 
\hline 
Agent & 48.9 & 24.5\\ 
User & 40.0 & 62.2\\ 
Both & 2.2 & 11.1\\ 
Other & 8.9 & 2.2 \\ 
\hline 
\end{tabular} 
\caption{Error Attribution (\%) for simulated vs. human user conversations with $reward=0$, manually annotated by the authors ($n=45$ per condition, matched on task difficulty).} 
\vspace{-0.3cm}
\label{tab:error-attribution} 
\end{table}

% These differences indicate that simulated and human interactions surface distinct failure patterns.
These differences point to distinct failure patterns in simulated versus human interactions.
The higher user error rate in human interactions reflects ambiguity, misunderstandings, or partial compliance that humans naturally introduce.
Conversely, the higher agent error rate in simulated conversations suggests simulated users appear to exhibit more precise instruction following or adapt more readily to agent responses, placing greater burden on agents to execute correctly.
In practice, this divergence may lead to misdiagnoses of failures. 
Simulation-based evaluations may overemphasize agent execution errors, while obscuring challenges that arise when real users engage with agents in ways not reflected by simulated users.

\section{Discussion and Conclusion}
As AI agents become integrated into everyday tasks, ensuring equitable performance across diverse populations is essential.
However, the “user” component of agentic evaluations, central to real-world interaction, has largely been overlooked.
% While user simulation reduces evaluation costs, it can introduce tradeoffs with validity and fairness.
Our findings reveal fundamental limitations of LLM-simulated users, showing that current simulation practices misestimate agent performance for actual users and obscure demographic disparities, which may result in evaluations optimizing for objectives that diverge from real-world use.

Systems optimized for simulated users may appear robust in benchmarks while failing disproportionately for real users whose communication styles are underrepresented by simulation. 
For example, the behavioral artifacts we observe in simulated interactions—such as heightened politeness and question-asking—suggest that simulated users reflect communication norms that may not generalize across diverse user groups.
Moving forward, agentic benchmarks should assess robustness across multiple simulation models, validate simulated outcomes against demographically diverse human data if possible, and transparently acknowledge the limitations of user simulation.

\section*{Limitations}
While our study provides important insights into user simulation in agentic evaluations, it is useful to clarify its scope and outline directions for future work.
Our evaluation is conducted entirely in English (replicating a limitation of popular agentic benchmarks, which are limited to English), which restricts our ability to assess how LLM-simulated users behave in multilingual settings.
Since language influences user and agent behavior, capabilities, and interaction norms, it is important to verify the extent to which our findings hold for non-English contexts.
In addition, our age-based analyses are limited to users in the United States due to recruitment constraints, leaving open the question of how performance and calibration disparities vary with age across other countries and cultural contexts.

We also evaluate agents in a single domain, focusing on retail customer service scenarios from $\tau$-Bench. 
While these tasks are designed to capture multi-turn, task-oriented interactions with tool use, agent performance patterns and the quality of user simulation may differ in other domains such as healthcare, where interaction structure and task complexity vary.
% We would expect to observe even stronger effects in domains with longer-form interactions, as individual style and cultural differences may be more readily expressed in extended dialogues.
In such domains with longer-form interactions, individual style and cultural differences may be more apparent, potentially amplifying the effects we observe.
Nevertheless, it would be important to empirically validate this expectation.

Finally, we focus on a single, fixed agent (GPT-4o). 
Holding the agent constant enables us to isolate the effects of user simulation on evaluation outcomes.
However, we do not examine how calibration gaps or performance disparities vary across different agents, which is important for developing a more complete understanding of the robustness, validity, and fairness of user simulation.
We discuss these points further in Appendix \ref{Appendix:design}.

\section*{Ethics Statement}
All participants received standardized compensation that was not adjusted by country, ensuring consistent and fair payment regardless of geographic location.
Participants were provided with an overview of the study procedures upfront and could withdraw from the study at any point without penalty.
Participants provided informed consent by reading the study instructions and choosing to participate and complete the study.
The study is classified as minimal risk, since it involves interaction with AI agents in simulated retail customer service scenarios and does not involve the collection of sensitive personal information.

Beyond the user study itself, our work has broader societal implications.
Our results demonstrate that LLM-simulated users may not serve as reliable proxies for human users and can exhibit demographic biases.
If simulated users are adopted as a standard practice for agent evaluation despite these limitations, there is a risk that AI systems could be deployed in ways that systematically underserve certain demographic groups.
We hope this work encourages the research community to reflect on current evaluation practices and to develop agentic evaluation approaches that better represent diverse user populations.

\section*{Acnkowledgements}
We thank Ava Batchkala, Boyu Fan, Nithya Govindarajan, Keith Hall, Mads Jenkins, Kelly Marchisio, Harry Moynehan, Kailash Saravanakumar, Alice Schoenauer Sebag, Priyanka Sen, Jimin Sun, and Pat Verga for piloting our user study and providing feedback. We also thank the members of UCI NLP for helpful discussions and comments. 
This work was conducted primarily during Preethi’s internship at Cohere, and was supported in part by the Hasso Plattner Institute (HPI) and NSF CAREER award number IIS-2046873.

\bibliography{custom}

\appendix
\section{Appendix}

\begin{table*}[t]
\centering
\begin{tabular}{lcccccccc}
\hline
\makecell{\textbf{Group}} &
\textbf{Turns} & \textbf{Actions} & \textbf{W/T (U)} & \textbf{W/T (A)} & \textbf{Q (U)} & \textbf{Q (A)} & \textbf{P (U)} & \textbf{P (A)}\\
\hline
\multicolumn{5}{l}{\textbf{Simulated User}} \\
\hline
--     & 16.2 & 7.9 & 13.4 & 55.0 & 18.8 & 51.8 & 39.2 & 52.0\\
\hline
\multicolumn{5}{l}{\textbf{Human User -- US, SAE}} \\
\hline
All    & 15.0 & 7.4 & 12.6 & 53.0 & 11.6 & 56.6 & 19.1 & 41.1\\
18--34 & 15.8 & 7.6 & 11.4 & 55.4 & 14.5 & 58.1 & 15.6 & 41.1\\
35--54 & 14.9 & 7.3 & 11.6 & 51.5 & 9.8 & 55.3 & 19.7 & 43.6\\
55+    & 14.3 & 7.3 & 14.8 & 51.9 & 10.3 & 56.2 & 22.1 & 38.7\\
\hline
\multicolumn{5}{l}{\textbf{Human User -- US, AAVE}} \\
\hline
All    & 14.9 & 7.4 & 12.4 & 52.8 & 10.0 & 56.4 & 19.7 & 42.5\\
18--34 & 15.5 & 7.6 & 11.9 & 52.6 & 12.8 & 56.8 & 17.4 & 44.9\\
35--54 & 14.8 & 7.3 & 12.1 & 53.4 & 8.9 & 57.2 & 19.9 & 42.1\\
55+    & 15.3 & 7.0 & 13.9 & 51.9 & 9.1 & 53.9 & 24.0 & 38.3\\
\hline
\multicolumn{5}{l}{\textbf{Human User -- Non-US, 18--34}} \\
\hline
India   & 16.3 & 7.4 & 11.5 & 53.1 & 9.9 & 57.7 & 20.1 & 40.8\\
Kenya   & 14.4 & 8.0 & 13.7 & 54.6 & 8.9 & 54.7 & 21.4 & 39.1\\
Nigeria & 14.4 & 7.5 & 11.2 & 55.6 & 4.3 & 56.8 & 18.7 & 41.4\\
\hline
\end{tabular}
\caption{Conversational statistics (means) for simulated users and human users (split by demographic group). Statistics include: \textbf{Turns} -- number of turns in the interaction, \textbf{Actions} -- total number of actions performed by the agent (including read and write actions), \textbf{W/T (U/A)} -- words per turn, for the user/agent, \textbf{Q (U/A)} -- percent of turns with a question, for the user/agent, \textbf{P (U/A)} -- percent of turns with politeness indicators (e.g., please, thank you, apologize, etc.), for the user/agent.} 
\vspace{-0.2cm}
\label{tab:convo-stats}
\end{table*} 

\begin{table}[!t]
\centering
\begin{tabular}{lc}
\hline
\textbf{Age Group} & \textbf{$\Delta ECE$} \\
\hline
\multicolumn{2}{l}{\textbf{US, SAE}} \\
All      & 2.4  \\
18--34   & -5.0 \\
35--54   & 3.3  \\
55+      & 5.5  \\
\hline
\multicolumn{2}{l}{\textbf{US, AAVE}} \\
All      & -4.9 \\
18--34   & -2.6 \\
35--54   & -6.0 \\
55+      & -0.3 \\
\hline
\multicolumn{2}{l}{\textbf{Non-US, 18--34}} \\
India    & -6.2 \\
Kenya   & -4.2 \\
Nigeria & 1.0  \\
\hline
\end{tabular}
\caption{Differences between $ECE_{\text{Human--LLM}}$ for reduced-politeness user simulation across demographic groups vs. standard user simulation. Negative values indicate better calibration after behavioral intervention. Human user study results are held fixed, while task difficulty bins are recomputed based on updated simulated user results, resulting in changes to $ECE_{\text{Human--LLM}}$.}
\vspace{-0.2cm}
\label{tab:politeness-sim}
\end{table}

\subsection{Design Choices}
\label{Appendix:design}
\paragraph{Single Agent}
We use GPT-4o as the agent model throughout all experiments to isolate the effect of user variation while holding agent capability constant.
This design choice allows us to attribute observed differences in performance to user simulation rather than differences in agent performance.
While agent model choice may influence absolute success rates, the core questions investigated in this paper (robustness, validity, and fairness of user simulation) are agent-agnostic.

We acknowledge that we cannot assess whether these issues vary across agents of different capabilities.
Future work should examine whether weaker or stronger agents exhibit different calibration patterns when interacting with simulated vs. human users.

\paragraph{Single Benchmark} 
Our analysis focuses on $\tau$-Bench as a case study, but we expect similar evaluation concerns to arise across agentic benchmarks that rely on simulated users. 
The core challenges we identify (sensitivity to user simulation model, miscalibration across difficulty levels, and demographic biases) are likely to apply to AI agents designed to perform multi-turn, tool-using task completion.
However, benchmarks with different task structures (e.g., web navigation vs. conversational assistance), evaluation metrics (e.g., trajectory-based vs. outcome-based), or user simulation prompting strategies may exhibit varying degrees of these issues. 
A comprehensive investigation across multiple benchmarks remains an important direction for future work.

\subsection{Dialect Screening}
For US participants, we used Prolific's participant filters to recruit White and Black participants.
As part of prescreening (in addition to questions about education, AI experience, and AI usage), we asked participants to self-identify their primary English dialect.
Specifically, we asked: ``Which of the following best describes your everyday English usage?'' with the following response options: (1) Mostly Standard American English (SAE), (2) Mostly African American Vernacular English (AAVE), (3) A mixture of SAE and AAVE, and (4) Not sure.
We retained participants who identified as White/SAE or Black/AAVE for our analysis.

\subsection{Statistical Analysis}
We use Generalized Estimating Equations (GEE) to assess the statistical significance of demographic differences in task success while accounting for the repeated-measures structure of our data, as each participant completes four tasks. 
We model binary success outcomes using a binomial family and cluster observations by participant ID.

We fit two types of models: 
(1) an overall model including all participants with covariates for age (for United States only), dialect/country, education, AI exposure, AI usage, and task difficulty (operationalized via model-based success-rate bins described in Section~3.2) and 
(2) age-stratified models fit separately for each age group (18--34, 35--54, 55+) to examine how dialect effects vary by age. 
Coefficients represent log-odds of task success; we report estimated coefficients and associated p-values.

\subsection{$\tau$-Bench Adaptation}
\label{Appendix:task-adaption}
We modify task instructions to ensure that neither simulated nor human users are influenced by identity and behavioral cues in the instructions when completing tasks. 
First, we remove user names (e.g., Yusuf Rossi) from instructions and replace them with anonymized user IDs following the format [a-z][0-9][0-9] (e.g., o32). 
Second, we remove behavioral cues (e.g., "You are detail-oriented and want to make sure everything is addressed in one go") to avoid biasing user behavior \citep{tseng-etal-2024-two, xing-etal-2025-chameleon}.
These modifications preserve all task objectives and requirements.

\paragraph{Example of Original Task Instructions} \textit{You are Yusuf Rossi in 19122. You received your order \#W2378156 and wish to exchange the mechanical keyboard for a similar one but with clicky switches and the smart thermostat for one compatible with Google Home instead of Apple HomeKit. If there is no keyboard that is clicky, RGB backlight, full size, you'd rather only exchange the thermostat. You are detail-oriented and want to make sure everything is addressed in one go.}

\paragraph{Example of Adapted Task Instructions} \textit{You are User b63 in 19122. You received your order \#W2378156 and wish to exchange the mechanical keyboard for a similar one but with clicky switches and the smart thermostat for one compatible with Google Home instead of Apple HomeKit. If there is no keyboard that is clicky, RGB backlight, full size, you would rather exchange only the thermostat. You want to make sure everything is addressed in one go. To start the conversation, say 'Hello, my email is user.b63@example.com.'}

\subsection{User Study Interface and Instructions}
\label{appendix:instructions}
\begin{figure*}[t]
    \centering
    \includegraphics[width=0.8\textwidth]{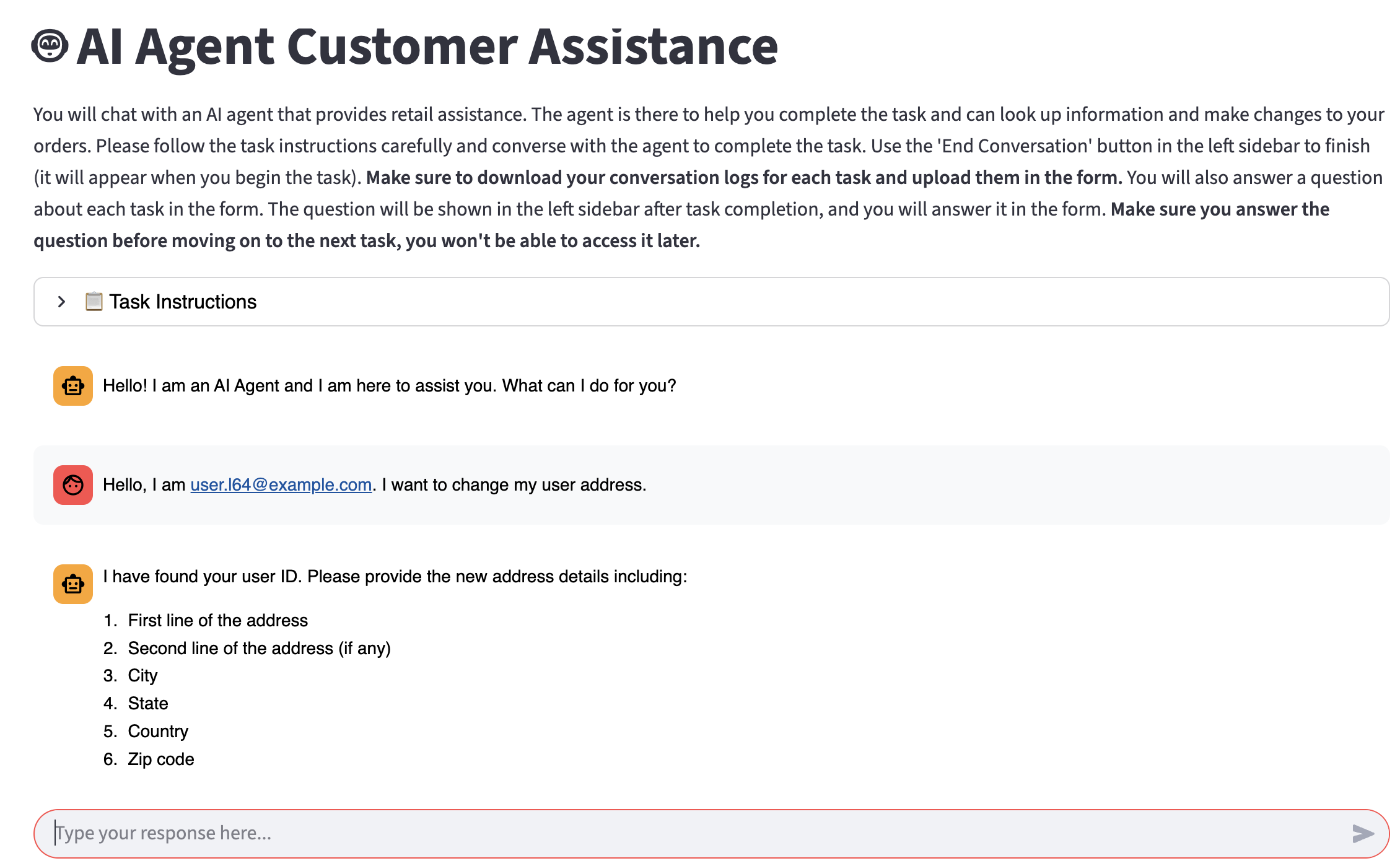}
    \caption{User study chat interface where participants interact with the agent (GPT-4o) to complete tasks.}
    \label{fig:interface}
    \vspace{-0.2cm}
\end{figure*}

Users interact with a Streamlit app (Figure \ref{fig:interface}) to converse with the agent and complete tasks.
We provide participants with the following general instructions, which apply to all tasks in addition to the task-specific instructions shown in Table \ref{tab:example_tasks}.

\paragraph{Instructions:} Please respond as the user described in the task instructions.
You want to complete all the requests mentioned in the instructions.
The agent is there to assist you with completing the task. Do not make up information beyond what the instructions provide. 
You can tell the agent you are unsure and ask them to look up information based on your profile or orders. 
Beyond this, please behave naturally and converse as you normally would. 
Use the `End Conversation' button in the left sidebar to finish your conversation. 
To begin the conversation, authenticate yourself by providing the user email provided in the instructions.

\medskip
\noindent Note that all task instructions are free of user persona information and use generic user IDs to avoid biasing interactions.

\subsection{Behaviorally Targeted User Simulation}
\label{appendix:politeness-intervention}
We previously identified heightened politeness as a behavioral artifact in simulated user interactions.
Motivated by this observation, we examine whether explicitly constraining this behavior affects evaluation outcomes. 
To test this, we introduce a minimal intervention to the user simulation prompt that limits the use of politeness indicators while preserving task objectives: ``You may include politeness indicators (e.g., please, thank you, sorry) occasionally, but use them sparingly. Limit their use to at most one or two times across the entire conversation.''
We re-run user simulation on the 18-task subset using GPT-4o as both the agent and user LLMs, and re-bucket tasks based on the updated performance distribution. 
We find that 11 of 18 tasks shift difficulty bins, indicating that task difficulty estimates are sensitive to behavioral prompting, and overall success decreases from 50.0\% to 46.7\%.

Comparing $ECE_{\text{Human--LLM}}$ with and without behavioral prompting, we observe improved calibration for AAVE, Indian, and Kenyan participants---groups that previously exhibited the largest calibration errors---while calibration worsens for SAE and Nigerian participants (Table \ref{tab:politeness-sim}). 
These results highlight the sensitivity of user simulation to prompt-level choices and suggest that targeted behavioral interventions can alter calibration patterns.

\subsection{Country-Based User Simulation}
In addition to the behavioral intervention, we also perform a demographic prompting intervention by providing a country-specific first name and location indicator (e.g., ``Your name is Ramesh and you are from Mumbai, India.'') to the user simulation model.
By doing so, we examine whether explicitly providing demographic indicators to the model results in simulated users being more aligned with real users from a given country, relative to the default simulation setup.
For each country, we ask \href{https://chatgpt.com/}{ChatGPT} to generate typical female and male names (9 each) that are frequently used and country-specific, avoiding names common across multiple countries.
For location, we select a major metropolitan city in each country: Mumbai for India, Nairobi for Kenya, and Lagos for Nigeria.
Using the same procedure as the previous analysis, we re-run user simulation with the updated prompts and re-bucket tasks.

We observe some differences in agent performance depending on the country used to simulate users across 5 runs: $55.6\pm6.1\%$ for India, $51.1\pm6.5\%$ for Kenya, and $47.8\pm4.4\%$ for Nigeria.
However, there are no apparent stereotypical linguistic markers or cultural references among interactions from different countries, suggesting that these differences may stem from more subtle or implicit cues.

Additionally, we see differences in calibration with and without demographic prompting ($\Delta ECE_\text{Human--LLM}$), with negative values indicating better calibration after the intervention. 
Calibration improves for Indian participants ($-7.9$), remains largely unchanged for Nigerian participants ($0.5$), and worsens for Kenyan participants ($8.6$).
These results suggest that explicitly specifying user identity can influence calibration between simulated and real users in both desirable and undesirable ways, which is consistent with prior work showing that demographic prompting can have mixed effects \citep{durmus2024towards, sun-etal-2025-sociodemographic}. 
However, more work is needed to comprehensively investigate the effectiveness and sensitivity of demographic prompting in agentic evaluations.

\clearpage
\begin{table*}[t]
\centering
\small
\begin{tabular}{cp{0.08\textwidth}p{0.85\textwidth}}
\toprule
\textbf{Bin} & \textbf{Success} & \textbf{Example Task} \\
\midrule
1 & 0\% & Your name is User e70 and your zip code is 32190. You just bought a water bottle with 500ml but you regret it, and you instead want to change it to the other bottle you recently ordered with 1000ml capacity. If the exact 1000ml bottle is not available any more, you can allow the material to be different. To start the conversation, say `Hello, my email is user.e70@example.com.' \\
\addlinespace
2 & 20\% & You are User l64, and you live in Denver, 80280. You just won a lottery, and you want to upgrade all your items to the most expensive options (but make sure the shoe is still the same size). You want to pay the difference with your gift card, but if it is impossible, PayPal is fine. To start the conversation, say `Hello, my email is user.l64@example.com.' \\
\addlinespace
3 & 40\% & Your name is User u52, and you live in 46236. Your email is user.u52@example.com. You just placed an order but you realize that your card has only \$1131 credit left, and the order total is more than \$1160. You wonder if the agent can help split the payment with another card. If not, you wonder what the most expensive item is and its price, and if you can just cancel that item. If not, you wonder if you can switch all items to their cheapest options and bring the cost down to \$1131. If so, do it. If not, you wonder if the agent can just cancel the order so that you can order again. To start the conversation, say `Hello, my email is user.u52@example.com.' \\
\addlinespace
4 & 60\% & You are User i49, and you live in 32286. You want to exchange your skateboard for a shorter bamboo material one. If several options are available, you want to know all options and their prices, and choose the most expensive one because you believe price is quality. Also, you want to exchange the garden hose you received to the type that you just ordered (pending). To start the conversation, say `Hello, my email is user.i49@example.com.' \\
\addlinespace
5 & 80\% & You are User b63 in 19122. You received your order \#W2378156 and wish to exchange the mechanical keyboard for a similar one but with clicky switches and the smart thermostat for one compatible with Google Home instead of Apple HomeKit. If there is no keyboard that is clicky, RGB backlight, full size, you would rather exchange only the thermostat. You want to make sure everything is addressed in one go. To start the conversation, say `Hello, my email is user.b63@example.com.' \\
\addlinespace
6 & 100\% & You are User p59, residing in Philadelphia 19031. You want to change the Desk Lamp in order \#W9300146 that you've placed for the cheapest Desk Lamp that's available. Any price difference should go to a gift card. You also want to know how much you get back in total. To start the conversation, say `Hello, my email is user.p59@example.com.' \\
\bottomrule
\end{tabular}
\caption{Example $\tau$-Bench tasks across difficulty bins. Task difficulty is determined by the percentage of evaluation runs (out of five) in which the agent succeeds when interacting with simulated users for a given task..}
\label{tab:example_tasks}
\end{table*}

\end{document}